\documentclass[aps,pra,reprint,showpacs,floatfix,superscriptaddress]{revtex4-1}

\usepackage{graphicx}
\usepackage{dcolumn}
\usepackage{bm}
\usepackage{xcolor}
\usepackage{amsmath}
\usepackage{multirow}
\begin{document}

\preprint{APS/123-QED}

\title{Quantum Metrology of Absorption and Gain Parameters\\ using Two-Mode Bright Squeezed Light} 

\author{Mrunal Kamble}
\affiliation{Department of Physics and Astronomy, Texas A\&M University, College Station, Texas 77843, USA}
\affiliation{Institute for Quantum Science and Engineering, Texas A\&M University, College Station, Texas 77843, USA}
\author{Jiaxuan Wang}
\affiliation{Department of Physics and Astronomy, Texas A\&M University, College Station, Texas 77843, USA}
\affiliation{Institute for Quantum Science and Engineering, Texas A\&M University, College Station, Texas 77843, USA}
\author{Girish S. Agarwal}
\affiliation{Department of Physics and Astronomy, Texas A\&M University, College Station, Texas 77843, USA}
\affiliation{Institute for Quantum Science and Engineering, Texas A\&M University, College Station, Texas 77843, USA}
\affiliation{Department of Biological and Agricultural Engineering, Texas A\&M University, College Station, Texas 77843, USA}





\begin{abstract}
Absorption and gain processes are fundamental to any light-matter interaction and a precise measurement of these parameters is important for various scientific and technological applications. Quantum probes, specifically the squeezed states have proved very successful, particularly in the applications that deal with phase shift and force measurements. In this paper, we focus on improving the sensitivity of the estimation of the photon loss coefficient of a weakly absorbing medium as well as the estimation of the gain parameter using a two-mode bright squeezed state. The generation of this state combines the advantage of a coherent beam for its large photon number with the quantum properties of the two-mode squeezing operation in an optical parametric amplifier. We present two measurement schemes: balanced photodetection and time-reversed metrology, both utilizing two-mode bright squeezed light. The maximum quantum advantage we can achieve using two-mode bright squeezed light is 3.7 times for the absorption parameter $\alpha = 0.05$ and 8.4 times for $\alpha = 0.01$ as compared to using only the coherent state. Similarly, the maximum quantum advantage for the estimation of optical gain is found around 2.81 times for the gain coefficient $G=1.05$ and around 6.28 times for $G=1.01$. We discuss the significance of using one measurement scheme over the other under different squeezing conditions. We compare our results with the Cramér-Rao bound for a two-mode bright squeezed state to assess the quality of the proposed methodologies. 

\end{abstract}

\maketitle


\section{\label{sec:level1} Introduction}   
Quantum sensing and metrology is one of the emerging fields that demonstrate the practical utility of quantum mechanical probes in various scenarios. The sensitivity of the optical systems using the coherent states of light is limited by the Standard Quantum limit (SQL). In the region where classical states of light are deemed insufficient to improve the measurement sensitivity, quantum states step in to push the boundaries \cite{PRXQuantum.3.010202,lawrie2019quantum}. Benefits of using quantum probes such as single-mode and two-mode squeezed states; and entangled states that surpass the SQL have been shown in phase shift measurements \cite{anderson2017phase, PhysRevLett.111.173601,caves1981quantum,jianqin}, force measurements \cite{pooser2020truncated,PhysRevA.87.012107}, spectroscopy and imaging\cite{li2022quantum,casacio2021quantum,pooser2016plasmonic,de2020quantum,triginer2020quantum}, magnetometry \cite{li2018quantum} to mention a few.


The standard optical interferometers like the Mach-Zehnder Interferometer play a crucial role in advancing quantum sensing capabilities by adding quantum probes as input to the interferometer \cite{plick,gerry2000heisenberg,pezze2008mach} as originally suggested by Caves \cite{caves1981quantum}. An important goal of such studies with different quantum probes was to find when the Heisenberg limit can be reached. Yurke, McCall, and Klauder \cite{yurke,zyou,caves2020reframing} introduced the SU(1,1) interferometer by replacing the beam splitters in the Mach-Zehnder interferometer with optical parametric amplifiers (OPA) and demonstrated the first possibility of reaching the Heisenberg limit of phase sensitivity.  This was an important step forward in quantum metrology and led to considerable experimental and theoretical work. A recent experiment \cite{jianqin} reports Heisenberg's limited measurement of phase using an SU(1,1) interferometer. More recently time-versed metrology is finding widespread applications with both single and two-mode squeezed vacuum states of light and even matter \cite{agarwal,vuletic,burd,linnemann,xin2019optimal,nanhuo}.

The work of Yurke et al. was generalized to add a seed beam to the OPA by Plick et al. \cite{plick}. This addition takes advantage of both the coherent component and the squeezed component in metrology. It leads to what is called the bright squeezed light and has been extensively used in studies on the quantum metrology of phase 
\cite{ou2012enhancement,yuhongliu2,anderson2017phase,liu2018quantum,PhysRevA.95.063843}. It also turned out that the SU(1,1) interferometer is more loss-tolerant and thus advantageous in applications \cite{santandrea2023lossy,tian2023loss,xin2021phase,lett,junxin,yuhongliu,manceau2017detection}.

While much of the work on sensing with quantum probes has been devoted to investigations on operations characterized by unitary transformations, studies of the parameters of the open systems are beginning to appear. The squeezed vacuum states have been shown to be especially useful \cite{gammaT,wang2024quantum,nair1,PhysRevLett.128.180506,moreau,losero,monras}. The two-mode squeezed vacuum states are optimum for the measurement of the absorption and gain parameters. While this is very important, it is so far difficult to produce squeezed vacuum states with a large number of photons and hence it is desirable to add a seed to obtain a large number of photons. This is the object of the present study. As mentioned in the abstract, absorption and gain are the fundamental processes the study of which yields a wealth of information on the molecular densities, transition matrix elements, molecular structure, etc. Thus absorption and gain spectroscopy techniques are very common tools and the sensitivities of such techniques can be improved considerably by using quantum light.


Here we explore a readily implementable entangled Gaussian state known as the Two-Mode Bright Squeezed State (TMBSS), where the squeezing operation is performed by a non-linear process in an optical parametric amplifier. This state combines the advantages of a coherent beam, with its high photon count, together with the noise reduction achieved through two-mode squeezing. It is to be noted that the goal of this study is to find the level of sensitivity using a good quantum resource rather than finding when the estimate of the parameter is optimum. In this study, we present results for quantum advantage in the measurements of both absorption and gain using TMBSS. 
The paper is structured as follows: In Section II, we introduce the characteristics of TMBSS. Section III delves into studying the Cramér-Rao bound of absorption estimation with TMBSS. In Section IV, we present measurement approaches for estimating the absorption parameter using TMBSS. Section V demonstrates the advantages of these methods by comparing them with the traditional approach employing a coherent light source. Our analysis encompasses the benefits provided by intensity differences measurement and the SU(1,1) interferometer under various squeezing conditions. Furthermore, we compare sensitivity results with the Cramér-Rao bound, derived in Sec. III. In Section VI, we present a detailed investigation of the gain using the Cramer-Rao bound and sensitivities for different types of measurements. Finally, Section VII gives a comprehensive discussion of our results.

\section{Two-Mode Bright Squeezed Light for Absorption Measurement}
In a two-mode bright squeezed state, a coherent field
and a vacuum field interacts non-linearly to produce a quantum state which is such that the difference in number fluctuations are squeezed. The Hamiltonian of the squeezing process is given by $H=\Gamma a^{\dagger}b^{\dagger}+h.c$ where $\Gamma$ indicates the interaction strength of this process. 
The squeezing operation is performed by a non-linear process in the Optical Parametric Amplifier (OPA) where the two input modes, a coherent state, and a vacuum state are acted upon by the two-mode squeezing operator given by
\begin{equation}
     {S}(r)=e^{r({a}^{\dagger} {b}^{\dagger}- {a} {b})},
\end{equation}
where $r$ is the squeezing parameter. We take the two-mode squeezing parameter to be real since absorption estimation does not consider any phase shifts. The resulting state called the two-mode bright squeezed state is given by
\begin{eqnarray}
    \left|\psi\right\rangle_{TMBSS} = {S}(r)\left|u,0\right\rangle,
\end{eqnarray}
where $\left|u\right\rangle $ is the coherent state such that $\left|u\right|^{2}\gg \cosh^{2}r$. The state generated by Eq. (2) is a Gaussian state. 

In the scheme shown in Fig.~1. the part with the OPA describes the generation of the TMBSS which is used to determine the parameter $\alpha$. It is more convenient to work with Heisenberg operators in this scenario. The squeezing transformations, $ {a_1}=  {S}^{\dagger} {a_{in}} {S}$ and $ {b_1}=  {S}^{\dagger} {b_{in}} {S}$ are given by
\begin{eqnarray}
 {a_{1}}&=& {a_{in}}\cosh r+ {b_{in}^{\dagger}}\sinh r, \nonumber\\
 {b_{1}}&=& {b_{in}}\cosh r+ {a_{in}^{\dagger}}\sinh r.
\end{eqnarray}
\begin{figure}
    \includegraphics[width=9.2cm,height=1.8cm, trim=0 0 0 0cm]{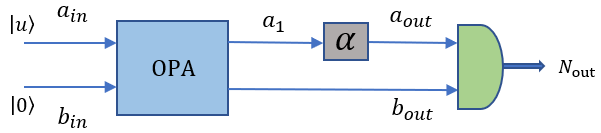}
    \caption{The schematics of absorption measurement using two-mode bright squeezed state. One of the input beams is the coherent beam and the other is the vacuum field. The two beams at the output of the OPA constitute the two-mode bright squeezed light. One of the beams called the probe beam passes through the absorption medium. Detection of the intensity difference of these beams $\left\langle{N}_{out}\right\rangle=\left\langle a_{out}^{\dagger}a_{out}\mathbin{-b_{out}^{\dagger}b_{out}}\right\rangle$ forms the signal for the balanced photodetection setup}
    \label{fig:enter-label}
\end{figure}
In the following step, one of the beams is directed into the sample and is called the probe, and the other beam is called the ancilla. After passing through the sample the probe carries the information on the absorption coefficient which is based on the beam splitter transformation model \cite{agarwal2012quantum} given by
\begin{eqnarray}
 {a_{out}}&=& {a_{1}}\sqrt{1-\alpha}+ {v}\sqrt{\alpha},\nonumber\\
 {b_{out}}&=& {b_{1}}.\label{absorption}
\end{eqnarray}
When working at the quantum level, we need to introduce the vacuum field given by annihilation operator ${v}$ to satisfy the commutation relation $\left[ {a_{1}}, {a}_{1}^{\dagger}\right]=\left[ {a_{out}}, {a}_{out}^{\dagger}\right]=1$. 
Using Eqs. (3) and (4) we can write
\begin{eqnarray}\label{output}
 {a}_{out}&=& {a_{in}}c_{11}+ {b_{in}^{\dagger}}c_{12}+ v\,c_{13},\nonumber\\
 {b}_{out}&=& {a_{in}^{\dagger}}c_{21}+ {b_{in}}c_{22}, 
\end{eqnarray}
where
\begin{gather}
    c_{11}=\sqrt{1-\alpha}\cosh r \nonumber, \hspace{0.1cm}c_{12}=\sqrt{1-\alpha}\sinh r\nonumber, \hspace{0.1cm}c_{13}=\sqrt{\alpha},\nonumber\\
    c_{21}=\sinh r\nonumber,\hspace{0.1cm} c_{22}=\cosh r\nonumber.
\end{gather}
 Lastly, the two beams enter the photocurrent mixer to realize the superposition of the fields which constitutes the signal for the measurement scheme discussed in section IV subsection A. 
 
 
\section{Quantum Fisher Information for Two-Mode Bright Squeezed state}
With the new emerging quantum technologies for sensing, it is important to know how far we can push the limits of precisely measuring a physical parameter. Quantum measurement theory provides us with this limit called the quantum Cram\'er-Rao bound  which is an extension of its classical counterpart. We drop the quantum prefix and denote it simply as the Cram\'er-Rao bound for the remainder of the paper. It is expressed in terms of the Quantum Fisher Information (QFI) of the given system that describes the limit on the distinguishability between two infinitesimally close quantum states. Typically, the larger the QFI, the better the distinguishability which implies better precision in estimating the parameter.

In the present research, the precision of the measurement is based on this Cram\'er-Rao bound. We first calculate the QFI to determine this bound. As described in the previous section the input fields to the OPA are Gaussian in nature. The general approach to computing the quantum Fisher information for a Gaussian system has been developed \cite{pinel2012ultimate, pinel2013quantum, gao2014bounds, friis2015heisenberg, vsafranek2015quantum, banchi2015quantum, marian2016quantum, nichols2018multiparameter, vsafranek2018estimation}. Given that the output field maintains the Gaussian characteristics from the absorbing medium \cite{agarwal1987wigner}, we can apply the covariance matrix method, which is well-formulated in \cite{vsafranek2018estimation}, to find the QFI associated with the TMBSS. The calculations are to be done using the explicit expressions for $ {a}_{out}$ and $ {b}_{out}$ given by Eq.~(5) and using the fact that $ {a}_{in}$ and $ {b}_{in}$ represent the coherent and the vacuum field respectively. We first write the set of the output annihilation and the creation operators for the given two-mode bosonic system in the form of a vector of operators given by
\begin{equation}
     {A}=(a_{out},b_{out},a_{out}^{\dagger},b_{out}^{\dagger})^{T}.
\end{equation}
Introducing the displacement vector $d_{m}=tr[\rho  {A}_{m}]$ and its covariance matrix $\sigma_{mn}=tr\left[ {\rho}\left\{ \Delta {A}_{m},\Delta {A}_{n}^{\dagger}\right\} \right]$ where $\Delta {A}= {A}-d$ and recognizing we only have one parameter to be estimated ($\alpha$), the QFI is given by
\begin{equation}
    F_{Q}(\alpha)=\frac{1}{2}vec\left(\frac{\partial\sigma}{\partial\alpha}\right)^{\dagger}\mathfrak{M^{-1}}vec\left(\frac{\partial\sigma}{\partial\alpha}\right)+2\frac{\partial d^{\dagger}}{\partial\alpha}\sigma^{-1}\frac{\partial d}{\partial\alpha}\label{qfi},
\end{equation}
where, $\mathfrak{M}=\overline{\sigma}\otimes\sigma-\mathit{K\otimes K}$ and $K=\left[\begin{array}{cc}
I & 0\\
0 & -I
\end{array}\right].$\newline
The required displacement vector and the covariance matrix can be calculated using Eq. (5) yielding the results:  
\begin{equation} 
    d^{(\alpha)}=\left(\begin{array}{c} u\cosh r\sqrt{1-\alpha}\\ u^{\ast}\sinh r\\ u^{\ast}\cosh r\sqrt{1-\alpha}\\ u\sinh r\end{array}\right)\label{d},
\end{equation}
and
\begin{equation}
    \sigma^{(\alpha)} = \begin{bmatrix}\label{s}
    \sigma_{11}^{(\alpha)} & 0 & 0 & \sigma_{14}^{(\alpha)}
    \\0 & \sigma_{22}^{(\alpha)} & \sigma_{23}^{(\alpha)} & 0
    \\0 & \sigma_{32}^{(\alpha)} & \sigma_{33}^{(\alpha)} & 0
    \\\sigma_{41}^{(\alpha)} & 0 & 0 & \sigma_{44}^{(\alpha)}
    \end{bmatrix},
\end{equation}
where
\begin{gather*}
    \sigma_{11}^{(\alpha)}=\sigma_{33}^{(\alpha)}=1+2(1-\alpha)sinh^{2}r,\nonumber\\
    \sigma_{22}^{(\alpha)}=\sigma_{44}^{(\alpha)}=1+2sinh^{2}r,\nonumber\\
    \sigma_{14}^{(\alpha)}=\sigma_{41}^{(\alpha)}=\sigma_{23}^{(\alpha)}=\sigma_{32}^{(\alpha)}=2\sqrt{1-\alpha}\cosh r\sinh r.\nonumber\\
\end{gather*}
Considering the limit $\left|u\right|^{2}\gg \cosh^{2}r$, we can drop the first term in Eq.~(\ref{qfi}), which leads to the simplified form given by
\begin{equation}
    F_{Q}(\alpha)=2\frac{\partial (d^{(\alpha)})^{\dagger}}{\partial\alpha}(\sigma^{(\alpha)})^{-1}\frac{\partial d^{(\alpha)}}{\partial\alpha}.\label{simpleQfi}
\end{equation}   
Thus, using Eqs. (\ref{d}-\ref{simpleQfi}), the QFI of a two-mode bright squeezed state wherein one of the modes undergoes absorption with a parameter $\alpha$ is expressed as follows:

\begin{equation}\label{fisheralpha}  
    {F}_Q(\alpha)=\frac{|u|^{2}\cosh^{2}r\cosh2r}{[1+\alpha(\cosh2r-1)](1-\alpha)}\,,
\end{equation}
where $|u|^{2}$ is the total number of photons in the input coherent state used to generate the two-mode bright squeezed beam and $r$ is the squeezing parameter. 
The Camer-Rao bound which is the lowest bound in the estimation of the given parameter is found by using the inequality
    
\begin{equation}
    (\Delta\alpha)_{CR}\geq\frac{1}{\sqrt{\mathcal{N}F_{Q}(\alpha)}},\label{crbound}
\end{equation}
where $F_{Q}(\alpha)$ is the quantum Fisher information Eq.~(\ref{fisheralpha}) and $\mathcal{N}$ is the number of independent measurements which we take as 1 for simplicity. 
Next, we discuss the practical schemes and assess the possibility of reaching the Cram\'er-Rao bound.

\section{Measurement Schemes}

\subsection{Balanced Photodetection}
 The sensitivity of measuring the absorption coefficient ($\alpha$) depends not only on the state of the light but also on the chosen measurement scheme. Here we calculate the sensitivity of the setup consisting of three main steps namely, squeezing, absorption, and detection as shown in Fig.~1. In section II we discussed the interaction between the probe beam and the absorption sample as formulated in Eq.~(\ref{absorption}). The ancilla does not interact with the sample and is used as a reference beam. In the next step, both the beams enter the balanced detector to realize the superposition of the fields. Here, the signal is defined as the intensity difference between the two beams such that, $\left\langle{N}_{out}\right\rangle=\left\langle a_{out}^{\dagger}a_{out}\mathbin{-b_{out}^{\dagger}b_{out}}\right\rangle $ and the associated fluctuations are given by $\Delta N_{out}=\sqrt{\left\langle N_{out}^{2}\right\rangle -\left\langle N_{out}\right\rangle ^{2}}$. Using Eq.~(5) these quantities are found to be

\begin{equation}
\left\langle{N}_{out}\right\rangle=|u|^{2}\left(c_{11}^{2}-c_{21}^{2}\right),
\end{equation}
and
\begin{eqnarray}
(\Delta N_{out})^2&=& |u|^{2}[c_{11}^{2}\left(c_{11}^{2}+c_{12}^{2}+c_{13}^{2}\right)+c_{21}^{2}\left(c_{21}^{2}+c_{22}^{2}\right)\nonumber\\
&-&2c_{11}c_{21}\left(c_{11}c_{21}+c_{12}c_{22}\right)].
\end{eqnarray}
The sensitivity of the setup is calculated using
\begin{equation}
    \Delta\alpha=\frac{\Delta N_{out}}{\mid d\left\langle{N}_{out}\right\rangle/d\alpha\mid}.\label{SenFormula}
\end{equation}
From Eqs.~(13-15) the sensitivity of the balanced photodetection measurement scheme could be written in terms of the squeezing parameter r and the absorption coefficient $\alpha$ as

\begin{equation}
    (\Delta\alpha)_{BD}=\frac{\sqrt{1+2\alpha^{2}\cosh^{2}r\sinh^{2}r-\alpha\cosh^{2}r}}{| u |\cosh^{2}r}.
\end{equation}
Next, we calculate the sensitivity of another popular method, namely, the SU(1,1) interferometer measurement scheme. 

\subsection{SU(1,1) Interferometer-Time reversed metrology}
\begin{figure}
\includegraphics[width=9.1cm,height=1.8cm, trim=-1 0 0 0cm]{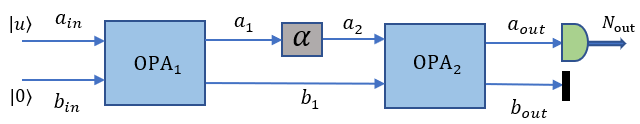}
\caption{Experimental SU(1,1) setup for bright squeezed state time-reversal metrology. The absorption medium, which reduces the output intensity to $(1-\alpha)$ times the input intensity, is placed between two optical parametric amplifiers (OPAs), on the upper arm of the interferometer. The first OPA generates bright squeezed light with coherent light at one of the input modes. The second OPA reverses the squeezing transformation. Detection of the number of outgoing photons $\left\langle a_{out}^{\dagger}a_{out}\right\rangle$ leads to estimating the photon-loss coefficient $\alpha$.}
\end{figure}


Here we present the SU(1,1) interferometer scheme which further improves the signal-to-noise ratio for absorption measurement at certain squeezing conditions in comparison with the balanced photodetection scheme. Referring to Fig.~2 this setup has one additional step before detection where the probe beam and the ancilla both enter another OPA where they undergo the anti-squeezing process. Thus, the second OPA reverses the squeezing transformation $r \rightarrow -r$ giving the new output states as
\begin{eqnarray}
 {a_{out}}&=& {a_{2}}\cosh r- {b_{1}^{\dagger}}\sinh r,\nonumber\\
 {b_{out}}&=& {b_{1}}\cosh r- {a_{2}^{\dagger}}\sinh r.
\end{eqnarray}
This additional step further improves the sensitivity of the setup as discussed ahead in the results section.
Using Eq.~(17) and replacing $ {a_{out}}$ in Eq.~(4) with $ {a_{2}}$ we can write the output operators as
\begin{eqnarray}
     {a}_{{\rm out}}=&& {a}_{{\rm in}}\left(\sqrt{1-\alpha}\cosh^{2}r-\sinh^{2}r\right)\nonumber\\&&
    - {b}_{{\rm in}}^{\dagger}\left(1-\sqrt{1-\alpha}\right)\sinh r\cosh r\nonumber\\&&
    + {v}\cosh r\sqrt{\alpha},\nonumber
\end{eqnarray}
\begin{eqnarray}
     {b}_{{\rm out}}=&& {b}_{{\rm in}}\left(\cosh^{2}r-\sqrt{1-\alpha}\sinh^{2}r\right)\nonumber\\&&
    + {a}_{{\rm in}}^{\dagger}\left(1-\sqrt{1-\alpha}\right)\sinh r\cosh r\nonumber\\&&
    - {v}^{\dagger}\sqrt{\alpha}\sinh r .
\end{eqnarray}
For simplicity, we write $ {a}_{out}$ and $ {b}_{out}$ as
\begin{eqnarray}
 {a}_{out}&=& {a_{in}}d_{11}+ {b_{in}^{\dagger}}d_{12}+ {v}d_{13},\nonumber\\
 {b}_{out}&=& {a_{in}^{\dagger}}d_{21}+ {b_{in}}d_{22}+ {v^\dagger}d_{23}.
\end{eqnarray}  
where
\begin{eqnarray}
    \begin{array}{cc}
         & d_{11}=\sqrt{1-\alpha}\cosh^{2}r-\sinh^{2}r, \\
         & d_{12}=-\left(1-\sqrt{1-\alpha}\right)\sinh r\cosh r\nonumber, d_{13}=\sqrt{\alpha}\cosh r\nonumber,\\
         & d_{21}=\left(1-\sqrt{1-\alpha}\right)\sinh r\cosh r, \\
         & d_{22}=\cosh^{2}r-\sqrt{1-\alpha}\sinh^{2}r\nonumber, d_{23}=-\sqrt{\alpha}\sinh r\nonumber.\\
    \end{array}
\end{eqnarray}
At this point, we need to extract the information on the absorption coefficient by detecting the mean number of photons and the corresponding variance. We first consider the case of $\langle{N}_{\rm out}\rangle =\langle a_{\rm out}^\dagger a_{\rm out}+ b_{\rm out}^\dagger b_{\rm out}\rangle$ which typically constitutes the signal for SU(1,1) technique. The signal and the fluctuations, in this case, are given by
\begin{equation}
    \left\langle N_{out}\right\rangle = |u|^{2}\left(d_{11}^{2}+d_{21}^{2}\right),
\end{equation} 
and,
\begin{eqnarray}
\Delta^{2}N_{out}&=&|u|^{2}[d_{11}^{2}\left(1+2d_{12}^{2}\right)+d_{21}^{2}\left(d_{21}^{2}+d_{22}^{2}+d_{23}^{2}\right)\nonumber\\
&+&2d_{11}d_{21}(d_{11}d_{21}+d_{12}d_{22}+d_{13}d_{23})],
\end{eqnarray}
respectively. The sensitivity in Eq.~(\ref{SenFormula}) applies to both the measurement schemes. Thus we get,
\begin{equation}
    (\Delta\alpha)_{+}=\frac{\sqrt{U}}{V},
\end{equation}
where
\begin{eqnarray}
    U&=&4-4\zeta-4\alpha\cosh^2r+2\zeta\cosh{4r}-2\alpha\eta\sinh^24r,\nonumber\\
    V&=&\frac{2\cosh^{2}r}{\sqrt{1-\alpha}}(2-\sqrt{1-\alpha}\cosh{2r}). \nonumber
\end{eqnarray}
We use $\zeta:=1-\sqrt{1-\alpha}$ and $\eta:=\alpha/4+\sqrt{1-\alpha}$ for simplicity. Notice that V in the denominator approaches zero when $r\rightarrow2$ for $\alpha=0.05$ and when $r\rightarrow3$ for $\alpha=0.01$. At these points $U \neq 0$, thus the sensitivity ratio has a singularity at $\sinh^{2}r=\frac{\sqrt{1-\alpha}}{2\zeta}$ as shown in Fig.~3(a). For small $\alpha$, $\sinh r\sim\frac{1}{\sqrt{\alpha}}$. 
Note that the singularity can be avoided by detecting the signal only from one of the ports such that $\langle{N}_{\rm out}\rangle=\langle a_{\rm out}^\dagger  a_{\rm out}\rangle$. In this case, we calculate the signal and the associated fluctuations as
\begin{equation}
    \left\langle{N}_{out}\right\rangle=|u|^{2}d_{11}^{2},
\end{equation}
\begin{equation}
    \Delta^{2}N_{out}=|u|^{2}d_{11}^{2}\left(1+2d_{12}^{2}\right).
\end{equation}
Using Eqs.~(15), (23-24) the sensitivity could be written in terms of r and $\alpha$ as 
\begin{equation}
    \Delta\alpha=\frac{\sqrt{(1-\alpha)(1+4(1-\sqrt{1-\alpha})^{2}\cosh^{2}r\sinh^{2}r)}}{| u|\cosh^{2}r}.
\end{equation}
Now that we have calculated the sensitivities from the measurement perspective, we can compare their quantum advantage and discuss the best possible schemes under different squeezing conditions. 


\section{Quantum Advantage: Sensitivity Enhancement in Measurement Schemes}
\begin{figure}[htbp]
    \hspace{-0.6cm}
    \includegraphics[width=8.5cm,height=10cm, trim=0 0 0 0cm]{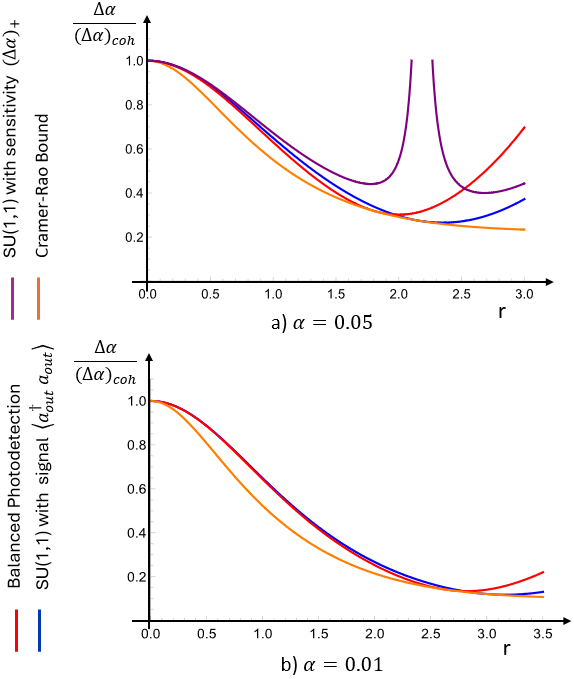}
    \caption{Sensitivities $\Delta\alpha/\Delta\alpha_{coh}$ (red line - balanced-photodetection, blue line - SU(1,1)) and the Cram\'er-Rao bound $(\Delta\alpha/\Delta\alpha_{coh})
    _{min}$ (orange line) using two-mode bright squeezed light, scaled by the sensitivity of using a coherent light without the interferometer, as a function of the squeezing parameter $r$, for weak absorption (a) $\alpha=0.05$ (b) $\alpha=0.01$. In addition, Fig.~3a also shows $\Delta\alpha_{+}$ (purple line).}
    \label{fig:enter-label}
\end{figure}

\begin{table}[]
    \centering
    \setlength{\tabcolsep}{0.4cm}
    \def\arraystretch{1.5}
    \begin{tabular}{|c|c|c|c|c|}
        \hline
        $\alpha$&r&$QA_{BD}$&$QA_{SU(1,1)}$&$QA_{CRB}$\\
        \hline
        \multirow{2}{*}{0.05} & 1.99 & \textbf{3.32} & 3.36 &3.42 \\
        \cline{2-5}
        & 2.35 & 2.81 & \textbf{3.77} & 3.85 \\
        \hline
        \multirow{2}{*}{0.01} & 2.82 & \textbf{7.39} & 7.53 & 7.63  \\
        \cline{2-5}
        & 3.17 & 6.26&\textbf{8.41}  &8.59 \\
        \hline
    \end{tabular}
    \caption{Quantum Advantage ($QA=\frac{\Delta\alpha_{coh}}{\Delta\alpha}$) for absorption measurement using: - balanced photodetection (BD), SU(1,1) interferometer, and the corresponding Cram\'er-Rao bound (CRB). Values displayed in bold are the best possible QA attainable using the specified measurement technique.}
    \label{tab: quantum advantage}
\end{table}
We scale the sensitivities derived using the measurement schemes with the sensitivity using only a coherent state of the same photon number entering the sample without the interferometer. The ratio for the balanced photodetection (BD) measurement scheme is expressed as
\begin{eqnarray}
    \begin{array}{cc}\label{a}
         & \left(\frac{\Delta\alpha}{\Delta\alpha_{coh}}\right)_{A}=\frac{\sqrt{(1+2\alpha^{2}\cosh^{2}r\sinh^{2}r-\alpha\cosh^{2}r)/(1-\alpha)}}{\cosh r} ,
    \end{array}
\end{eqnarray}
and, the ratio for SU(1,1) interferometer with signal $\langle a_{\rm out}^\dagger a_{\rm out}\rangle$ is written as
\begin{eqnarray}
    \begin{array}{cc}\label{b}
         &\left(\frac{\Delta\alpha}{\Delta\alpha_{coh}}\right)_{B}=\frac{\sqrt{1+4(1-\sqrt{1-\alpha})^{2}\cosh^{2}r\sinh^{2}r}}{\cosh r}
    \end{array},
\end{eqnarray}
where the sensitivity using only a coherent state without an interferometer is given by
    \begin{equation}
        \Delta\alpha_{coh}=\frac{\sqrt{1-\alpha}}{|u|\cosh r}.
    \end{equation}
Fig.~3 shows the plots of the scaled sensitivities against the squeezing parameter r for (a) $\alpha=0.05$ and (b) $\alpha=0.01$. Eq.~(\ref{a}) and and Eq.~(\ref{b}) are represented by the red line and the blue line respectively. The Cra\'mer Rao bound Eq.~(\ref{crbound}) is represented by the orange line. As displayed in Table~I, both the setups can attain a quantum advantage ($QA=\frac{\Delta\alpha_{coh}}{\Delta\alpha}$) of greater than 3 for $\alpha=0.05$ and a quantum advantage of greater than 7 for $\alpha=0.01$. For $\alpha=0.05$ maximum advantage for the BD technique is achieved at $r=1.99\,(QA=3.32)$, and for the SU(1,1) technique is achieved at  $r=2.35\,(QA=3.77)$. For $\alpha=0.01$ maximum advantage for the BD technique is achieved at $r=2.82\,(QA=7.39)$ and for the SU(1,1) technique is achieved at  $r=3.17\,(QA=8.41)$. This result is significant in the weak absorption domain. As we calculate the sensitivity of the measurement it is worth noting the significance of the large number of photons in the two-mode bright squeezed state given the $1/\sqrt{N}$ dependence, where N is the total number of photons passing through the sample at a given time. For example, around $r=2$, $N$ is approximately $14|u|^2$, where $|u|^2$ is the total number of photons in the input coherent field. 

Looking at the plots in Fig.~3, we find out that the balanced photodetection method gives a slightly greater quantum advantage relative to the SU(1,1) method just before it reaches its maximum around $r=2$ for $\alpha=0.05$ and $r=2.82$ for $\alpha=0.01$. However, using the full SU(1,1) scheme can provide a higher quantum advantage for $r\geq2$ for $\alpha=0.05$ and $r\geq2.82$ for $\alpha=0.01$. Consequently, it might be advantageous to use the balanced photodetection method given its benefit of less number of optical components leading to fewer experimental errors. However, SU(1,1) interferometer can prove beneficial when technological advancements allow for the higher squeezing parameters $r\geq2$. 

Next, the scaled sensitivity of both measurement schemes is compared to the theoretical limit given by the Cram\'er-Rao bound (orange line) which is obtained by taking the ratio of sensitivity Eq.~(\ref{crbound}) calculated using the quantum Fisher information of the two-mode bright squeezed state and that for the coherent state again given by $\Delta\alpha_{coh}=\frac{\sqrt{1-\alpha}}{|u|\cosh r}$. Note that the QFI of the TMBSS remains unaffected by the anti-squeezing action of the second OPA. Our results exhibit a close correspondence with the Cram\'er-Rao bound, specifically in the region between $1.5\leq r\leq2.5$ for $\alpha=0.05$ and $2\leq r\leq3$ for $\alpha=0.01$. This emphasizes the sensor's exceptional performance as measured against the theoretical lower bound for the given system. 

\section{Two-Mode Bright Squeezed Light for Gain Measurement: Cram\'er-Rao Bound and Measurement Schemes}

Apart from absorption, gain is another fundamental physical phenomenon that has broad applications in sensing and imaging technology. For example, gain sensing proves valuable in detecting Unruh-Hawking radiation using single-mode probes \cite{PhysRevLett.105.151301}.
Typically, the gain parameter associated with these processes is small, as it relies on population inversion within the medium. 
Optical amplification mechanisms such as stimulated emission in LASER media, Stimulated Raman Scattering (SRS), and Stimulated Brillouin Scattering (SBS) are characterized by this gain parameter. 
The gain obtained from SBS, for instance, serves as a signal for measuring the viscoelastic properties of the medium in imaging and spectroscopic applications. 
Similarly, the gain from SRS is utilized to characterize the nonlinear properties of optical fibers, which are critical components in telecommunication systems. 
Like absorption, the gain measurement is sensitive to system noise, necessitating advanced techniques to achieve precise measurement \cite{xu2022quantum, triginer2020quantum}.
Nair et al. \cite{PhysRevLett.128.180506} have considered optimal measurement of gain using quantum probes. However, the advantages offered by adding a coherent seed while generating the two-mode squeezed state remain unexplored. Thus here, we again consider the Two-Mode Bright Squeezed State (TMBSS) for estimating the gain parameter. 
We first calculate QFI for gain measurement using the TMBSS and provide the best possible sensitivity for the gain parameter which is the Cram\'er-Rao bound. Considering Fig.~1 but for the gain medium, the annihilation operators after the sample written in terms of the input operators are given by, 
\begin{eqnarray}
    \begin{array}{cc}
         &  {a_{out}}= {a_{1}}\sqrt{G}+ {v}^{\dagger}\sqrt{G-1}, \\
         &  {b_{out}}= {b_{1}}.
    \end{array}    
\end{eqnarray}

\begin{figure}[htbp]
    \hspace{-0.0cm}
    \includegraphics[width=8cm,height=5.2cm, trim=0 0 0 0cm]{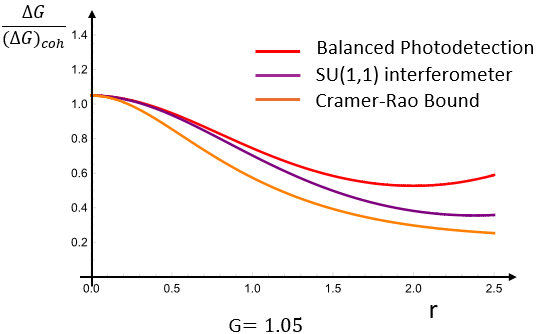}
    \caption{Sensitivities $\Delta G/\Delta G_{coh}$ (red line - balanced photodetection, purple line - SU(1,1) with intensity addition) and the Cram\'er-Rao Bound $\Delta G_{CRB}/\Delta G_{coh}$ (orange line) using two-mode bright squeezed light, scaled by the sensitivity of using a coherent light without the interferometer, as a function of the squeezing parameter $r$, for gain $G=1.05$.}
    \label{fig:enter-label}
\end{figure}
\noindent The Gaussian characteristics of the input state are preserved under the transformation in a gain medium \cite{agarwal1987wigner}. Thus, following the same process as used in Section III, 
the displacement vector and the covariance matrix are:  
\begin{equation} 
    s^{(G)}=\left(\begin{array}{c} u\sqrt{G}\cosh r\\ u^{\ast}\sinh r\\ u^{\ast}\sqrt{G}\cosh r\\ u\sinh r\end{array}\right)\label{d2},
\end{equation}
\noindent and
\begin{equation}
    \beta^{(G)} = \begin{bmatrix}\label{s2}
    \beta_{11}^{(G)} & 0 & 0 & \beta_{14}^{(G)}
    \\0 & \beta_{22}^{(G)} & \beta_{23}^{(G)} & 0
    \\0 & \beta_{32}^{(G)} & \beta_{33}^{(G)} &0
    \\\beta_{41}^{(G)} & 0 & 0 & \beta_{44}^{(G)}
    \end{bmatrix},
\end{equation}
where
\begin{gather*}
\beta_{11}^{(G)}=\beta_{33}^{(G)}=2G\cosh^{2}r-1,\nonumber\\
\beta_{22}^{(G)}=\beta_{44}^{(G)}=1+2sinh^{2}r,\nonumber\\
\beta_{14}^{(G)}=\beta_{41}^{(G)}=\beta_{23}^{(G)}=\beta_{32}^{(G)}=2\sqrt{G}\cosh r\sinh r.\nonumber\\
\end{gather*}
We obtain the QFI as 
\begin{equation}
    F_{Q}(G)=\frac{|u|^{2}\cosh^{2}r\cosh2r}{G[G+(G-1)\cosh2r]},
\end{equation}
which leads to $(\Delta G)_{CR} = [F_{Q}(G)]^{-1/2}$. Considering the measurement schemes, we perform similar calculations as done previously to determine the sensitivity of measuring the gain parameter using the two measurement schemes. For the balanced photodetection method, the output modes are given by
\begin{eqnarray}   {a}_{out}&=& {a_{in}}m_{11}+ {b_{in}^{\dagger}}m_{12}+ {v^\dagger}m_{13},\nonumber\\
 {b}_{out}&=& {a_{in}^{\dagger}}m_{21}+ {b_{in}}m_{22},   
\end{eqnarray}\label{output}  
where,
\begin{gather}
    m_{11}=\sqrt{G}\cosh r\nonumber, \hspace{0.1cm}m_{12}=\sqrt{G}\sinh r\nonumber, \hspace{0.1cm}m_{13}=\sqrt{G-1},\nonumber\\
    m_{21}=\sinh r\nonumber,\hspace{0.1cm} m_{22}=\cosh r.\nonumber
\end{gather}
Using the intensity difference $I_d$ between two beams as the signal we calculate the sensitivity for this setup to be
\begin{equation}
    (\Delta G)_{BD}=\frac{\sqrt{I_d(2I_d-|u|^2)}}{|u|\cosh ^2 r},
\end{equation}
where $I_d=|u|^2(G\cosh ^2 r-\sinh ^2 r)$.

For the SU(1,1) interferometer scheme with gain medium, the final output operators are given by
\begin{eqnarray}
 {a}_{out}&=& {a_{in}}f_{11}+ {b_{in}^{\dagger}}f_{12}+ {v^\dagger}f_{13},\nonumber\\
 {b}_{out}&=& {a_{in}^{\dagger}}f_{21}+ {b_{in}}f_{22}+ {v}f_{23},
\end{eqnarray}  
where
\begin{eqnarray}
    \begin{array}{cc}
         & f_{11}=\sqrt{G}\cosh^{2}r-\sinh^{2}r, \\
         & f_{12}=\left(\sqrt{G}-1\right)\sinh r\cosh r\nonumber, f_{13}=\sqrt{G-1}\cosh r\nonumber,\\
         & f_{21}=-\left(\sqrt{G}-1\right)\sinh r\cosh r, \\
         & f_{22}=\cosh^{2}r-\sqrt{G}\sinh^{2}r\nonumber, f_{23}=-\sqrt{G-1}\sinh r\nonumber.\\
    \end{array}
\end{eqnarray}
Considering $\langle{N}_{\rm out}\rangle =\langle  a_{\rm out}^\dagger  a_{\rm out}+  b_{\rm out}^\dagger  b_{\rm out}\rangle$ as the signal and calculating the corresponding fluctuations $\Delta N_{out}$, we find the sensitivity as
\begin{equation}
    (\Delta G)_+=\frac{\sqrt{GA}}{B},
\end{equation}
where
\begin{eqnarray*}
A &=& \mu(7G^{3/2}-3G+5\sqrt{G}-1) \\
&+&4(3\sqrt{G}\mu-\nu)\mu\nu\cosh2r \\
&+&8\mu^{2}\nu^{2}\cosh4r+4\mu\nu^{3}\cosh6r+\nu^{4}\cosh8r), \\
B&=&4\cosh^{2}r(1+\nu\cosh2r).
\end{eqnarray*}
We use $\mu:=\sqrt{G}+1$ and $\nu:=\sqrt{G}-1$ for simplicity.

In Fig.~4 we plot the scaled sensitivities ($\Delta G/\Delta G_{coh}$) for both the measurement 
schemes along with the Cram\'er-Rao bound $(\Delta G)_{CR}/\Delta G_{coh}$ where
\begin{equation}
    \Delta G_{coh}=\frac{\sqrt{G}}{|u|\cosh r}.
\end{equation}
Here, we find out that the SU(1,1) method gives a greater quantum advantage relative to the balanced photodetection method for any given squeezing parameter $r$. The best possible quantum advantages for both measurement schemes are shown in Table II. For $G=1.05$ at $r=2.3$ SU(1,1) is 2.81 times and the balanced detection is 1.78 times better, and for $G=1.01$ at $r=3.17$, SU(1,1) is 6.28 times and the balanced detection is 3.93 times better than using a coherent source. We can also see that both methods, specifically the SU(1,1) interferometer closely follow the Cram\'er-Rao bound which is the theoretical lower limit for the given scheme of measurements. 
    \begin{table}[]
    \centering
    \setlength{\tabcolsep}{0.4cm}
    \def\arraystretch{1.5}
    \begin{tabular}{|c|c|c|c|c|}
        \hline
        G & r & $QA_{BD}$ & $QA_{SU(1,1)}$ & $QA_{CRB}$ \\
        \hline
        1.05 & 2.37 & 1.78 & 2.81 & 3.82 \\
        \hline
        1.01 & 3.17  & 3.93 & 6.28& 8.58\\
        \hline
    \end{tabular}
    \caption{Quantum advantage for the gain measurement schemes: - balanced photodetection (BD), SU(1,1) interferometer, and the corresponding Cram\'er-Rao bound (CRB)}
    \label{tab: quantum advantage}
\end{table}
\section{Discussion}
This paper investigates the quantum advantage of using a Two-Mode Bright Squeezed State (TMBSS) to estimate the absorption parameter in a weakly absorbing medium as well as the small gain parameter in any optical amplification process. We present two setups – A) balanced photodetection and B) SU(1,1) interferometer, and discuss the best possible measurement schemes under different squeezing conditions. The paper's main goal is to showcase the improvement in the sensitivity measurement of the absorption or the gain parameter of the sample placed in one of the beam paths generated from an optical parametric amplifier. This study combines the benefits of a large number of photons in a coherent state with the quantum properties of the two-mode squeezing operation.

We first calculate the Quantum Fisher Information (QFI) of the two-mode bright squeezed state which leads us to the theoretical limit called the Cram\'er-Rao bound for measuring the absorption sensitivity. We then calculate the absorption sensitivities of both measurement schemes. We scale both these sensitivities by using only the coherent source with the same number of photons but without the interferometer.

The results show these setups can attain more than 7-fold Quantum Advantage (QA) for $\alpha=0.01$  and 3-fold QA for $\alpha=0.05$. This is an important result for the weak absorption domain. Plotting the scaled sensitivities against $r$ reveals the significance of using one measurement scheme over the other in different squeezing conditions. For example, the balanced photodetection method gives a slightly greater quantum advantage relative to the SU(1,1) method just before it reaches its maximum around r = 2 for $\alpha=0.05$. Thus, given the current technological limitations on the squeezing parameter $r$, the balanced photodetection method is beneficial because of its comparatively smaller number of optical components minimizing the experimental errors. However, the SU(1,1) interferometer gives a higher quantum advantage as we attain $r\geq2$ beyond which the QA for balanced detection goes on decreasing. Thus, the SU(1,1) interferometer holds promise for future absorption-related applications, particularly as technological progress enables higher squeezing parameters $r$.  

For the gain measurement, we find out that the SU(1,1) method shows a better quantum advantage than the balanced photodetection method for any given squeezing parameter $r$. For $G=1.05$ at $r = 2.37$, SU(1,1) is 2.81 times and the balanced detection is 1.78 times better than just using the coherent source without any interferometer. Comparing the scaled sensitivity of both the measurement schemes with the Cram\'er-Rao bound confirms the quantum advantage of the proposed setups in the estimation of absorption as well as the gain parameters.   

\section{ACKNOWLEDGMENTS}
The authors are grateful for the support of the Air Force Office of Scientific Research (Award No. FA-9550-20-1-0366), the Robert A. Welch Foundation (A-1943-20210327), and the Department of Energy (Grant No. FWP-ERW7011).


\bibliographystyle{ieeetr}
\bibliography{manuscript}

\end{document}